\documentstyle[12pt,citews]{article}

\newcommand{\bcn}{\begin{center}}
\newcommand{\beq}{\begin{equation}}
\newcommand{\beqn}{\begin{eqnarray}}
\newcommand{\ecn}{\end{center}}
\newcommand{\eeq}{\end{equation}}
\newcommand{\eeqn}{\end{eqnarray}}

\newcommand{\sect}[2]{\vspace*{6mm}\hspace*{-\parindent}{\bf #1.}~{\bf
#2}\vspace*{4mm}}

 \def\lsim{\mathrel{\rlap{\lower4pt\hbox{\hskip1pt$\sim$}}
    \raise1pt\hbox{$<$}}}
\def\slash#1{\setbox0=\hbox{$#1$}#1\hskip-\wd0\hbox to\wd0{\hss\sl/\/\hss}}
\begin{document}
\rightline {KU-HEP-93-30}

\bcn
{\bf $q\bar q$ bound states in the Bethe-Salpeter formalism}
\vspace*{1cm}

{\bf Pankaj Jain and Herman J. Munczek}\\

{\it Department of Physics and Astronomy\\
     University of Kansas\\
     Lawrence, KS-66045-2151\\}\vspace*{1cm}

{\bf ABSTRACT}
\ecn
\noindent
We solve the Bethe-Salpeter equation in order to calculate the
spectrum of pseudoscalar, vector and scalar meson bound states
for light as well as heavy quarks, extending our previous calculation
of pseudoscalar mesons. The fermion propagators appearing in the
equation are
obtained by solving the Schwinger-Dyson equation consistently with
the Bethe-Salpeter equation, a procedure necessary for demonstrating the
Goldstone nature of the pion in the chiral limit.
 We probe in our calculations a
model for the gluon propagator which leads to the expected three
dimensional potential in configuration space, while maintaining
its known ultraviolet behavior. Our procedure
provides a unified description of both the light and heavy mesons
and obeys known relations derived on the basis of current
algebra for light quarks and nonrelativistic analysis for heavy
quarks. The results
are in good agreement with experiment for the mass spectrum and
for the pseudoscalar mesons' decay constants. We also calculate
electromagnetic and isospin mass splittings for these mesons.

\bigskip

\noindent
PACS numbers: 12.38.Lg, 11.10.St, 11.30.Qc, 13.20.-v
\vfill
\eject
\hspace*{3pc}

\sect{1}{Introduction}

In a recent publication \citenum{JM92} we used a relativistically
covariant formalism to determine the
spectrum and properties of pseudoscalar mesons. The basic idea was to
solve the Bethe-Salpeter (BS) equation consistently with the Schwinger-
Dyson (SD) ladder equation for the quark propagator, a procedure necessary to
demonstrate the Goldstone nature of the pion in the chiral limit
\citenum{NJL,MISC1}. The
gluon propagator needed for the calculation exhibited the leading
ultraviolet behavior known from renormalization group analysis and its
infrared functional form was chosen such that it gave a harmonic
oscillator potential at intermediate distances
in three dimensional configuration space, a form which
is known from non-relativistic potential models \citenum{FLSC}
to lead to a realistic mass-spectrum for heavy mesons. The
resulting BS equation was solved by expanding the wave function
in terms of Tschebyshev polynomials, and keeping only the leading
 order functions. The subleading contributions were estimated numerically
to be small.

The numerical treatment  of the BS and SD equations presented in Ref. [1]
has the remarkable characteristic that it is equally applicable over all the
 spectrum of $q\bar q$ pseudoscalar bound states. There is no need to
resort to a Chiral-Lagrangian phenomenological limit for light mesons
or to nonrelativistic approximations for heavy mesons. The ultra-violet
asymptotic numerical behavior of the quark-mass functions and bound-state
wave functions agrees with the behavior derived from arguments
involving operator-product-expansion and renormalization-group
considerations \citenum{POL,MIR}.
 Results for the light $q\bar q$ ground state masses
were found to be in good qualitative and quantitative agreement
with current algebra results \citenum{GOR},
 while it was found that for heavy
quarks such as b and c the masses of their bound states are
proportional to the sum of their "constituent" masses in accordance
with nonrelativistic limit expectations. The decay constants $f_\pi$,
$f_K$ and those of the other pseudoscalars were predicted to
have values compatible with experimental bounds and with alternative
calculations, as well as with nonrelativistic limit predictions for
light-quark-heavy-quark bound states.

In the present paper we extend the methods of Ref. [1] to include
vector and scalar mesons. We also calculate their electromagnetic and
isospin mass-splittings as well as those of the pseudoscalar mesons.
The purpose of this extension is twofold. First, such a calculation
tests the predictive power of the methods used and of the
assumptions made about the
gluon propagator. Second, a knowledge of the bound state wave functions
for pseudoscalar, scalar and vector mesons is necessary for the detailed
calculation of a variety of processes such as meson leptonic,
semi-leptonic and non-leptonic decays, form factors, etc.

The procedure used in the present paper follows the general guidelines
of Ref. \citenum{JM92}, but is considerably simpler, more systematic
and accurate. The basic simplification arises from the assumption
that the dominant contribution comes from just one tensor component
for a particle of given spin and parity. We further assume that the first
term in the Tschebyshev expansion of this wave function gives the
dominant contribution. These assumptions were then explicitly
checked by calculating the corrections due to all the next to
leading order wave functions. For pseudoscalar and vector particles
these corrections were found to be small justifying our procedure.
 For the case of scalar mesons a truncation to
just one wave function was not found to be adequate. However a simple
extension of the above procedure
in which two, rather than one, wave functions are
included self consistently was found to be sufficient. The corrections
due to the remaining wave functions were also calculated and
found to be small. The procedure can be systematically improved to include
the corrections due to further dropped pieces.

In Section 2 we give the basic formulas and definitions. We also discuss
the approximation procedures involved in our calculations. Section 3
presents the discussion of our results for
the mass spectrum, the pseudoscalar leptonic
decay constants and the electromagnetic mass splittings.
 In the Appendix we give the detailed form of
the bound state equations.

\bigskip
\sect{2}{Basic Formalism}

The BS equation for a $q\bar q$ bound state of mass $m^2_{ab} = p^2$
 can be written in the ladder approximation as,

\beq
S_a^{-1}(q+\alpha p)\chi(p,q)S_b^{-1}(q-\beta p) = -i\int\gamma_\mu
\chi(p,k)\gamma_\nu G_{\mu\nu}(k-q){d^4k\over (2\pi)^4}\ ,
\eeq

with $\alpha+\beta = 1$.
The quark propagators are determined by solving the SD equation
for a fermion flavor $a$, also in the ladder approximation,
\beq
 S_a^{-1}(q) = \slash q - \tilde m_a + i\int\gamma_\mu S_a(k)
\gamma_\nu G_{\mu\nu}(k-q){d^4k\over (2\pi)^4}\ ,
\eeq
where $S^{-1}(q)$ can be written as,
\beq
S^{-1}(q) = \slash q\ A(q^2)\ -\ B(q^2)\ .
\eeq

It has been argued \citenum{LAND1,LAND2} that the ladder approximation is
most reliable in the Landau gauge,
which is adopted in the present calculation as it was in Ref. \citenum{JM92}.
In this gauge, the  gluon propagator $G_{\mu\nu}(k^2)$ appearing in
 equations (1,2) has the form.
\beq
G_{\mu\nu}(k) = -\bigg(g_{\mu\nu} - {k_\mu k_\nu\over k^2}\bigg) G(k^2)
\eeq

 In Euclidean variables we write
$G(k^2)$ as,

\beq
G(k^2) = {16\pi\over 3}\bigg[ {\pi d\over k^2\ln (x_0+x)}\bigg( 1+b
{\ln\lbrack\ln(x_0+x)\rbrack\over \ln(x_0+x)}\bigg)\bigg]
+ G_{IR}(k^2)
\eeq

The first term in this equation represents the
known ultraviolet asymptotic,
or large $k^2$, behavior known from renormalization group analysis.
For $x_0=0$, the term in square
brackets gives the two loop asymptotic expression
for the running coupling $\alpha_s(k^2)$,
where $d=12/(33-2n),\ \ b = 2\beta_2/\beta_1^2,\ \ \beta_1 = n/3-11/2,
\ \ \beta_2 = 19n/12-51/4$, and $n$ is the number of quark flavors,
which we set to five for our calculations. We also define $x=k^2/\Lambda_{QCD}
^2$. $\Lambda_{QCD}$ is the QCD scale parameter
which we take here to be 228  MeV. We have
introduced the parameter $x_0$ to regulate the infrared behavior of the
running coupling. As long as its value is larger than about 2, the
results of the calculations are largely insensitive to it. As in Ref. [1]
we adopt a value $x_0=10$. The infrared effects are dominated by $G_{IR}
(k^2)$, which we choose to be of the form,

\beq
G_{IR}(k^2) = {16\pi^2\over 3}a\ k^2\ e^{-\mu k^2}\ ,
\eeq

\bigskip
with $a=(0.387\ GeV)^{-4}$ and $\mu=(0.510\ GeV)^{-2}$. In this case Eq. (5)
corresponds in three dimensional configuration space to the
potential given in Figure 1. As can be seen from the figure the
potential is roughly linear up to a distance of about 1 fermi.

The most general decomposition for the vector, pseudoscalar and
scalar bound state wave functions is given by the following expressions

{\it Vector:}
\beqn
\chi_{_V}(p,q) &=& \slash \epsilon \chi_{_{V0}} + \slash p \epsilon\cdot q
\chi_{_{V1}} + \slash q\epsilon\cdot q \chi_{_{V2}} + \epsilon\cdot q
\chi_{_{V3}} \nonumber\\
&+& [\slash\epsilon,\slash p]
\chi_{_{V4}} + [\slash\epsilon,\slash q]\chi_{_{V5}} +
[\slash p,\slash q]\epsilon\cdot q
\chi_{_{V6}} + i\gamma_5\slash t\chi_{_{V7}}
\eeqn
with
$t_\mu = \epsilon_{\mu\nu\alpha\beta}q_\nu p_\alpha \epsilon_\beta$,
 $\epsilon^2 = -1$ and $ \epsilon\cdot p = 0$.

{\it Pseudoscalar:}
\beq
\chi_{_P}(p,q) = \gamma_5\bigg[\chi_{_{P0}} + \slash p\chi_{_{P1}} +
\slash q\chi_{_{P2}}
+ [\slash p,\slash q]\chi_{_{P3}}\bigg]
\eeq

{\it Scalar:}
\beq
\chi_{_S}(p,q) = \chi_{_{S0}} + \slash p\chi_{_{S1}} + \slash q\chi_{_{S2}}
+ [\slash p,\slash q]\chi_{_{S3}}
\eeq

\bigskip
The flavor indices are left implicit in the $\chi_{_{Vi}}$,
$\chi_{_{Pi}}$ and $\chi_{_{Si}}$ which are
 functions of $p^2$, $q^2$ and
$p\cdot q$. Following the procedure of Ref. \citenum{JM92} we assume that
a Wick rotation to Euclidean variables is allowed for both the
SD and BS equations. Alternatively, we can derive the SD and BS
equation from the Euclidean path integral formulation of the theory,
thus avoiding possible difficulties in performing the Wick
rotation \citenum{WICK}. We can then write $p\cdot q=pq\cos\theta$
and expand the functions $\chi_{_{Ji}}$
in terms of Tschebyshev polynomials
$T^{(n)}(\cos\theta)$ \citenum{MRFS}:

\beq
\chi_{_{Ji}}\biggl(q^2,M_B^2,\cos
\theta\biggr)=\sum\limits_n\chi_{_{Ji}}^{(n)}\biggl(q^2,M_B^2\biggr)T^{(n)}
(\cos\theta)\; ,
\eeq
\bigskip
where the subindices $J$ and $i$ indicate either $V,\ P$ or $S$ and
1, 2, etc., respectively.
As we discuss later the dominant contributions to our calculation
for vector and pseudoscalar mesons come
from the first term in the expansion of $\chi_{_{J0}}$, namely
from $\chi_{_{J0}}^{(0)}$.
The corrections due to the
next two sub-leading terms in the Tschebyshev expansion as well as those
due to the other tensor components $\chi_{_{Ji}}^{(0)}$ (i=1,2, etc.)
 were calculated to be small, but were also included in our
final results.

\medskip
\sect{2a} {Leading order analysis and its corrections}
\medskip

The complete BS equation for the dominant
wave functions $\chi_{_{J0}}$ can be written for Euclidean
$p$ and $q$, as
\bigskip

{\noindent {\it Vector meson:}}

\beq
\chi_{_{V0}}(p^2,q^2,p\cdot q) =
 {1\over A_aA_bD}\lbrace (q^2-\alpha\beta p^2+m_am_b) +
q\cdot p (\alpha - \beta)\rbrace I_{V0}\ +\Delta\chi_{_{V0}},
\eeq
where,
\beq
I_{V0} = \int {d^4k \over (2\pi)^4} G(k-q)\chi_{_{V0}}\bigg\lbrace 1
+{2\over 3}
\bigg(1-{q\cdot (k-q)^2\over q^2(k-q)^2}\bigg)\bigg\rbrace
\eeq

{\noindent  {\it Pseudoscalar meson:}}
\beq
\chi_{_{P0}}(p^2,q^2,p\cdot q) =
{3\over A_aA_bD}\lbrace (q^2-\alpha\beta p^2+m_am_b) +
 (\alpha - \beta) q\cdot p\rbrace I_{P0}+\Delta\chi_{_{P0}}\ ,
\eeq
where,
\beq
I_{P0} = \int {d^4k \over (2\pi)^4} G(k-q)\chi_{_{P0}}
\eeq

{\noindent {\it Scalar meson:}}
\beq
\chi_{_{S0}}(p^2,q^2,p\cdot q) =
{3\over A_aA_bD}\lbrace (q^2-\alpha\beta p^2-m_am_b) +
 (\alpha - \beta) q\cdot p\rbrace I_{S0}+\Delta\chi_{_{S0}}\ ,
\eeq
where,
\beq
I_{S0} = \int {d^4k \over (2\pi)^4} G(k-q)\chi_{_{S0}}
\eeq

In the above expressions we have used the following notation,

\beq
A_a=A_a(q+\alpha p)\ \ \ \ \ \ A_b=A_b(q-\beta p)\ , \nonumber
\eeq
\beq
m_a = {B_a(q+\alpha p)\over A_a(q+\alpha p)}\ \ \ \
m_b = {B_b(q-\beta p)\over A_b(q-\beta p)}\ ,\nonumber
\eeq
and
\beq
D = [(q+\alpha p)^2+m_a^2][(q-\beta p^2)+m_b^2]\ .
\eeq

The $\Delta\chi_{_{J0}}$ contain contributions
 from several of the wave functions $\chi_{_{Ji}}$.
We expand the wave functions in terms of Tschebyshev polynomials
and project the equations with the order zero Tschebyshev
polynomial. Only $\chi_{_{J0}}^{(0)}$ will appear on the
left hand side but the right hand side will in general
get contributions
from all orders. The right hand side is approximated by keeping only
the leading order term in the Tschebyshev expansion for
the $\chi_{_{Ji}}$ for all $i$ greater than zero and by including
the three lowest terms, namely $\chi_{_{J0}}^{(0)}$,
$\chi_{_{J0}}^{(1)}$ and $\chi_{_{J0}}^{(2)}$ for $\chi_{_{J0}}$.
For the vector and pseudoscalar mesons we express the functions
$\chi_{Ji}^{(0)}$, $i$ larger than zero, and $\chi_{_{J0}}^{(1)}$ and
$\chi_{_{J0}}^{(2)}$
in terms of of their dependence
on just $\chi_{_{J0}}^{(0)}$ as given by equations (32) to (40)
and (42) to (46) in the Appendix.
In this
way we obtain an equation which contains only $\chi_{_{J0}}^{(0)}$ as
displayed by Eqns. (20) and (22).
For the scalar mesons, however, the contribution due to $\chi_{_{S2}}^{(0)}$
was found to be nonnegligible. We therefore solved the equation
for $\chi_{_{S2}}^{(0)}$ self consistently with the equation for
$\chi_{_{S0}}^{(0)}$. The corrections due to the remaining
wave functions were calculated by expressing them in terms of their
dependence on just  $\chi_{_{S0}}^{(0)}$. The resulting equations
are given by (24) and (47) to (52) in Appendix A.

In arriving at the equations given in the Appendix
 we expanded the mass functions and
the functions $A$ and $B$ in Taylor series around $q^2+\alpha^2p^2$ or
$q^2+\beta^2p^2$, depending on their argument, and retained up to
first derivative terms which typically contributed corrections of less than
10-15\% to the calculated meson masses.
The functions defined in Eqs. (17-19) were obtained by solving the
SD equation for space like values of their arguments and extrapolated
numerically to time like values.
The projected BS equations were then solved numerically for
bound state mass eigenvalues
and wave functions by
using standard routines.

\bigskip
\sect{2b}{Electromagnetic and isospin mass splittings}
\medskip

There has been considerable work, based on a variety of approaches,
 on the calculation of the
electromagnetic and isospin mass splittings \citenum{EMIS}.
In our formalism,
the electromagnetic effects can be easily incorporated
by including the one photon exchange kernel to the BS
equation.
To be consistent with the Goldstone's theorem in the chiral symmetric limit
 it was also necessary to include the electromagnetic
self energy corrections to the SD equation.

The isospin mass splittings were also included by using different
bare masses for the up and down quarks.
 The combined isospin and electromagnetic mass splittings
give the total mass differences between states of the type $q\bar u$
and $q\bar d$ where $q$ can be any of the quarks.
For the case of the  pion, however,
we assumed
that the purely isospin mass difference between $\pi^0$ and $\pi^+$ is
negligible.

\bigskip
\sect{3}{Numerical results and discussion}
\medskip

We first give a qualitative discussion of the numerical results.
The mass functions were obtained by solving the Schwinger-Dyson
equation with results basically the same as in Ref. [1]. The
quantitative differences are small since the gluon propagator is not
drastically changed. Fig. 2 shows the mass functions.
As in Ref. \citenum{JM92} our results for the light pseudoscalar mesons satisfy
the relations derived by using current algebra and for the heavy
mesons follow the behavior expected from nonrelativistic  considerations.
In comparison to Ref. \citenum {JM92},
in the present paper we have followed a simplified but more
accurate approach
for the calculation of
the bound state spectrum. The
success of the present approach is crucially dependent on the
relative magnitude of the corrections due to the subleading
wave functions. By explicit calculation
we have found that the corrections to the leading order result due
to higher order Tschebyshev polynomials and subleading tensor
components given by the $\Delta\chi_{_{J0}}^{(0)}$ displayed in
the Appendix amount to less than 10\% of the calculated meson masses.
 This gives us confidence  that we have isolated the
dominant contributions. We found that the
relative correction gets smaller as the mass of the meson increases.
However as the meson mass increases, the effective coupling decreases and
 a small correction to the mass
may mean a large correction to the binding energy. For very heavy
mesons, for example toponium,
 the nonrelativistic approximation  may prove to be a more practical
approach than the present covariant procedure. However, for even
heavier quarks with masses of the order of 500 GeV, the Higgs
interaction gets very large and again the present covariant procedure may
be more reliable \citenum{JSM92}.

The calculated mass functions are plotted in Fig. 2. The cutoff
dependent bare mass values chosen for the present calculation
are $\tilde m_u(\Lambda^2) = 3.1$ MeV, $\tilde m_s(\Lambda^2) = 73$ MeV,
$\tilde m_c(\Lambda^2) = 680$ MeV, $\tilde m_b(\Lambda^2) = 3.3$ GeV.
The cutoff $\Lambda$ is equal to 134 GeV. The current algebra
masses for the up and strange quark, defined in equation (3.1)
of Ref. [1] are given by $\hat m_u = 8.73$ MeV, $\hat m_s = 203$ MeV.
For the heavy quark a more physical quantity is the constituent
mass which we may define as the mass at the origin and is given by
$m_c$(0) = 1.54 GeV, $m_b$(0) = 4.74 GeV.
The up quark mass given above represent an average of the
up and down quark masses. The difference between the up and down
quark masses is given by $\hat m_d-\hat m_u$ = 4.33 MeV.

The results
for the ground state meson spectrum are given in Tables 1, 2 and 3.
Table 4 gives some calculated excited states and Table 5 gives
the results for the electromagnetic and isospin mass splittings.
Fig. 3,4 and 5 give
plots for some of the wave functions.
 The variation of the
masses with the center of mass parameter
$\alpha$ was found to be quite
small for low lying mesons and larger for heavier mesons.
For example, the pion mass changed by only about 3\% as $\alpha$ was varied
from 0.5 to 0.1 while the $\rho$ meson mass changed by about 14 \% as
$\alpha$ varied from 0.5 to 0.25.
The final values of $\alpha$, as quoted in Table 1, 2 and 3
 were chosen by the principle of minimal sensitivity in
the same fashion as discussed in Ref. [1]. In a few cases no extremum
was found but still there was a range of momentum values for
which the sensitivity was minimal. For these cases the quoted
value refers to the value roughly in the middle of this range.
  For the pseudoscalar decay
constants we followed the same procedure as in Ref. \citenum{JM92}.
{}From the tables it can be seen that, except for the $\pi$ and
$\rho$ mesons, the calculated hyperfine splitting is somewhat
larger than the one observed experimentally. The size of the
splitting varied with the values of the parameters $a$ and $\mu$
appearing in equation (6) for the gluon propagator, but not to the
degree needed. On the other hand, the splitting between the
radially excited
states and ground state is in general good
 agreement with experiment. We included in Table 4
only those excited states for which we felt that the extrapolations
of the functions $m(q^2)$ and $A(q^2)$ to timelike (negative) $q^2$ values were
numerically sound.  An interesting feature of the mass
functions is that while $m_u(q^2)$ is lower than $m_s(q^2)$ near the
space like region, the extrapolation gives a crossing with $m_u$
becoming larger than $m_s$. This explains the anomalous value of
$\alpha$ for $K^*$ as compared to that for $K$,
shown on the tables.

The above discussion and the tables 1-4 show that the present
procedure is sound and beginning to yield reasonable results
with very few adjustable parameters: the five quark masses
$m_a(\Lambda^2)$ and the constants $a$ and $\mu$ (as we mentioned
earlier the results are quite insensitive to $x_0$).
There remain, however, two
difficulties which require further study. One of the problems is
the larger than experimental
splitting between vector and pseudoscalar mesons for almost
all the states except for the $\pi$, $\rho$ system. We have checked
that this problem remains for all the parameter values and a few
different models that we studied for the gluon propagator.
This may signal the need for some effective scalar
confining interaction \citenum{ISGR}. In the present framework
this might be achieved by going beyond the ladder approximation
with the inclusion of multi-gluon exchange contribution. We
hope to study such contributions in a further investigation.

Another question
that we encountered and that requires further scrutiny
 is related to the extrapolation of the mass
functions into the time like region. This extrapolation was made
by fitting a fifth order polynomial to the low momentum region
of the mass functions. The extrapolation is
reasonable as long as the results are insensitive to higher order
contributions in this polynomial fit.
This was indeed found to be true for almost all of the
ground states of particular spin and parity but for some of the
radial excitations this extrapolation was found not reliable.
In the tables only those values are quoted for which the
extrapolation was found to be reasonable.

\bigskip

\sect{Acknowledgements}

We thank Douglas W. McKay and John Ralston for useful discussions.
This work has been supported in part by
DOE grant No. DE-FG02-85-ER-40214.A008.

\bigskip
\sect{Appendix A} {Bethe-Salpeter Equations}

In this appendix we give explicitly the formulas for the Bethe-Salpeter
equation treated in the approximation discussed in section 2. All
dimensionful variables have been scaled by $\Lambda_{QCD}$.
\bigskip

{\noindent {\it Vector meson equation:}}

\beq
\chi_{_{V0}}^{(0)}(p^2,x) =
\bigg\lbrace (x-\alpha\beta p^2+m_am_b)J_1 +
(\alpha -\beta + 2\alpha m'_am_b-2\beta m'_bm_a)J_2\bigg\rbrace I_{V0}
+\Delta\chi^{(0)}_{_{V0}}\ ,
\eeq

where,
\beq
I_{V0} = {2\over 3\pi}
\int dyy  \chi_{_{V0}}^{(0)}\bigg\lbrace K_1
+{2y\over 3}K_3\bigg\rbrace\ ,
\eeq

and $m'$ means the derivative with respect to the argument.
\bigskip

{\noindent {\it Pseudoscalar meson equation:}}

\beq
\chi_{_{P0}}^{(0)}(p^2,x) =
3\bigg\lbrace (x-\alpha\beta p^2+m_am_b)J_1 +
(\alpha -\beta + 2\alpha m'_am_b-2\beta m'_bm_a)J_2\bigg\rbrace I_{P0}
+\Delta\chi^{(0)}_{_{P0}}\ ,
\eeq

where,
\beq
I_{P0} = {2\over 3\pi}\int dyy \chi_{_{P0}}^{(0)}K_1\ .
\eeq

\bigskip

{\noindent {\it Scalar meson equation:}}

\beq
\chi_{_{S0}}^{(0)}(p^2,x) =
3\bigg\lbrace (x-\alpha\beta p^2-m_am_b)J_1 +
(\alpha -\beta - 2\alpha m'_am_b+2\beta m'_bm_a)J_2\bigg\rbrace I_{S0}
+\Delta\chi^{(0)}_{_{S0}}\ ,
\eeq

where,
\beq
I_{S0} = {2\over 3\pi}\int dyy \chi_{_{S0}}^{(0)}K_1\ .
\eeq

In the equations above we have

\beqn
J_1 &=& {2\over \pi}\int\limits^\pi_0 d\theta{\sin^2\theta \over
D(p^2,x,\cos\theta)}\nonumber\\
&=& {2 \over A_aA_b(\beta c_1c_4+\alpha c_2c_3)}
\bigg\lbrace{\alpha c_2\over D_1} + {\beta c_4\over D_2}
+ {d_1\over A_aA_b}\bigg\lbrack{c_1\over D_1}-{c_3\over D_2}\bigg\rbrack
\bigg\rbrace\ ,
\eeqn

\beqn
 J_2 &=& {2p\sqrt x\over \pi}\int\limits^\pi_0 d\theta{\sin^2\theta \over
D(p^2,x,\cos\theta)} \cos\theta\nonumber\\
&=& {1\over A_aA_b}{1\over (\alpha c_2c_3+\beta c_1c_4)}
\bigg\lbrace {c_3\over D_2} - {c_1\over D_1}\bigg\rbrace\nonumber\\
&+& {d_1\over 2\alpha\beta (A_aA_b)^2}{1\over (\alpha c_3+\beta c_1)}
\bigg\lbrace -{2\beta c_1^2\over D_1} - {2\alpha c_3^2\over D_2}
+\beta c_1 + \alpha c_3\bigg\rbrace\ ,
\eeqn

\beqn
D(p^2,x,\cos\theta)
& =& A_aA_bc_1c_3 + p\sqrt x \cos\theta\lbrack 2d_1c_1c_3+2\alpha c_2c_3A_aA_b
-2\beta c_1c_4A_aA_b \rbrack\nonumber\\
& + & p^2x\cos^2\theta\lbrack -4\alpha\beta c_2c_4A_aA_b+4\alpha d_1c_3 -
4\beta d_1c_1\rbrack \nonumber\\
&-& 8\alpha\beta p^3x^{3/2}\cos^3\theta\ d_1
\eeqn

 $$c_1 = x + \alpha^2p^2 + m_a^2$$
$$c_2 = 1 + m_am'_a$$
$$c_3 = x + \beta^2p^2 + m_b^2$$
$$c_4 = 1 + 2m_bm'_b$$
$$d_1 = \alpha A'_aA_b - \beta A'_bA_a,$$

\beq
D_1 = c_1+\sqrt{c_1^2-4\alpha^2 p^2 x c_2^2}
\eeq

\beq
D_2 = c_3+\sqrt{c_3^2-4\beta^2 p^2 x c_4^2}
\eeq

$\theta$ is the angle between $p$ and $q$, $x=q^2/\Lambda_{QCD}$,
$y=k^2/\Lambda_{QCD}$
$m_a$, $m'_a$, $A_a$ and $A'_a$ are functions of $x+\alpha^2p^2$ and
$m_b$, $m'_b$, $A_b$ and $A'_b$ are functions of $x+\beta^2p^2$\ .

\bigskip

The $\Delta\chi_{_{J0}}^{(0)}$, which represent corrections to
the leading order results, are given by the expressions below.

\bigskip

{\noindent {\it Vector mesons}:}

\beqn
\Delta\chi_{_{V0}}^{(0)} &=&{2\over 9\pi}
\bigg\lbrack -(x-\alpha\beta p^2+m_am_b)J_1+(\alpha-\beta)
J_2\bigg\rbrack \int dy y
\chi_{_{V2}}^{(0)}[2yK_7+y(y-x)K_3]\nonumber\\
&-&{4\over 9\pi}\bigg\lbrack (x-\alpha\beta p^2+m_am_b)J_2+xp^2(\alpha-\beta)
J_3\bigg\rbrack\int dyy\chi_{_{V1}}^{(0)}
{y\over \sqrt x}(\sqrt y K_4-\sqrt xK_3)\nonumber\\
&+&{2\over 3\pi}\bigg\lbrack 2p^2(\alpha m_b+\beta m_a)J_1 -
2(m_a-m_b)J_2\bigg\rbrack\nonumber\\
&\times & \int dyy \bigg\lbrack \chi_{_{V4}}^{(0)}(K_1-{4y\over 3}K_3)
+{\chi_{_{V6}}^{(0)}\over 3}(2yK_7+y(y-x)K_3)
\bigg\rbrack\nonumber\\
&+& {2\over 3\pi}\bigg\lbrack -2x(m_a-m_b)J_1+2(\alpha m_b+\beta m_a)J_2
\bigg\rbrack
\int dyy\chi_{_{V5}}^{(0)}\sqrt{y\over x}(-K_6+2\sqrt{xy}K_3)\nonumber\\
&+& {2\over 3\pi}\bigg\lbrack
-2x(m_a-m_b)J_2+2p^2x(\alpha m_b+\beta m_a)J_3\bigg\rbrack\nonumber\\
&\times & \int dyy\bigg\lbrack-{2\chi_{_{V4}}^{(0)}\over 3x}
(3K_1-4yK_3)+{2\chi_{_{V6}}^{(0)}y\over 3\sqrt x}
(\sqrt yK_4-\sqrt xK_3)\bigg\rbrack\nonumber\\
&+& xp^2(J_1-J_3){2\over 3\pi}\int dyy\chi_{_{V7}}^{(0)}\bigg\lbrack
-\sqrt{y\over x}-{2\over 3}yK_3)\bigg\rbrack\nonumber\\
& - & {2\over 3\pi}
\lbrack 2(x-\alpha\beta p^2+m_am_b)J_2 + 2xp^2J_3(\alpha-\beta)\rbrack
\ \int dyy\chi^{(1)}_{_{V0}}\sqrt{y\over x}\lbrack K_6+{2\over 3}
yK_4\rbrack\nonumber\\
& - & {2\over 3\pi}\ p^2
\lbrack (x-\alpha\beta p^2+m_am_b)J_1 + (\alpha-\beta)J_2\rbrack\nonumber\\
&\times &
\int dyy^2\chi^{(2)}_{_{V0}}\lbrack -K_1+{4\over 3}K_7-{2y\over 3}K_3+
{8y\over 15}K_8\rbrack\nonumber\\
& - & {8\over 9\pi}\ p^2
\lbrack (x-\alpha\beta p^2+m_am_b)J_3 + (\alpha-\beta)J_4\rbrack\nonumber\\
& \times & \int dyy^2\chi^{(2)}_{_{V0}}\lbrack 3K_1-4K_7+2yK_3-
{12y\over 5}K_8\rbrack\ ,
\eeqn

where
\beqn
\chi^{(0)}_{_{V1}} &= & {(\alpha-\beta)J_1\over (x-\alpha\beta p^2+m_am_b)J_1+
(\alpha-\beta)J_2}\ \chi_{_{V0}}^{(0)}\nonumber\\
&-& \bigg\lbrack {2x(\alpha-\beta)\over \alpha m_b+\beta m_a}
\  - {4J_2\alpha \beta\over J_1(\alpha m_b+\beta m_a)}
\bigg\rbrack\ \ \chi^{(0)}_{_{V6}}\ ,
\eeqn

\beq
\chi^{(0)}_{_{V2}} = {2J_1\over (x-\alpha\beta p^2+m_am_b)J_1+
(\alpha-\beta)J_2}
\chi_{_{V0}}^{(0)}+
{-x-\alpha\beta p^2+m_am_b\over \alpha m_b+\beta m_a}\ 2\chi^{(0)}_{_{V6}}\ ,
\eeq

\beqn
\chi^{(0)}_{_{V3}} &= & {(m_a+m_b)J_1\over (x-\alpha\beta p^2+m_am_b)J_1+
(\alpha-\beta)J_2}\ \chi_{_{V0}}^{(0)}\nonumber\\
&-& \bigg\lbrack {2x(m_a+m_b)\over \alpha m_b+\beta m_a}
\  + {2J_2(\alpha m_b-\beta m_a)\over J_1(\alpha m_b+\beta m_a)}
\bigg\rbrack\ \ \chi^{(0)}_{_{V6}}\ ,
\eeqn

\beq
\chi^{(0)}_{_{V4}} = -{1\over 2}\ {J_1(\beta m_a+\alpha m_b)\over J_1(x-
\alpha\beta p^2+m_am_b) +J_2(\alpha-\beta)}\ \chi^{(0)}_{_{V0}}\ ,
\eeq

\beq
\chi^{(0)}_{_{V5}} = {1\over 2}
{J_1(m_a-m_b)\over J_1(x-\alpha\beta p^2+m_am_b)+J_2(\alpha-\beta)}
\ \chi^{(0)}_{_{V0}}\ ,
\eeq

\beq
\chi^{(0)}_{_{V6}} = -{2\over 9\pi}
{\alpha m_b+\beta m_a\over x}J_1\ \int dyy\chi^{(0)}_{_{V0}}\ \bigg\lbrack
3K_1(x,y)-4yK_3(x,y) \bigg\rbrack\ ,
\eeq

\beq
\chi^{(0)}_{_{V7}} = {J_1\over J_1(x-\alpha\beta p^2+m_am_b)+J_2(\alpha-\beta)}
\ \chi^{(0)}_{_{V0}} \ ,
\eeq

\beq
\chi^{(1)}_{_{V0}} = {(2J_2/xp^2)\ (x-\alpha\beta p^2+m_am_b) + 2J_3
(\alpha-\beta)\over
J_1(x-\alpha\beta p^2+m_am_b)+J_2(\alpha-\beta)}
\ \chi^{(0)}_{_{V0}} \ ,
\eeq

\beq
\chi^{(2)}_{_{V0}} = {(4J_3-J_1)\ (x-\alpha\beta p^2+m_am_b) + (4J_4-J_2)
(\alpha-\beta)\over
J_1(x-\alpha\beta p^2+m_am_b)+J_2(\alpha-\beta)}
\ \chi^{(0)}_{_{V0}} \ .
\eeq

\bigskip

{\noindent {\it Pseudoscalar mesons}:}

\beqn
\Delta\chi_{_{P0}}^{(0)} &=&
\bigg\lbrack -{2\over 3\pi} (\beta m_a+\alpha m_b)J_1p^2  \int dy y
\chi_{_{P1}}^{(0)}(K_1+2yK_3)\nonumber\\
&+&{2\over 3\pi}(m_a-m_b)J_2
\int dyy\chi_{_{P1}}^{(0)}(3K_1-2 yK_3)\nonumber\\
&-&{4\over 9\pi}p^2(\alpha m_b+\beta m_a)J_3
\int dyy \chi_{_{P1}}^{(0)}(3K_1-4yK_3)\bigg\rbrack\nonumber\\
&+& {2\over 3\pi}\bigg\lbrack (m_a-m_b)J_1-(\alpha m_b+\beta m_a)
{J_2\over x}\bigg\rbrack
\int dyy\chi_{_{P2}}^{(0)}\sqrt{yx}(3K_6-2\sqrt{xy}K_3)\nonumber\\
&+& {2\over 3\pi}p^2 (J_1-J_3)
 \int dyy \chi_{_{P3}}^{(0)}
\bigg\lbrack(2\sqrt{xy}K_6-{8\over 3}xyK_3)\bigg\rbrack\nonumber\\
&+& {2\over \pi}\bigg\lbrack (x-\alpha\beta p^2+m_am_b)2J_2
+(\alpha-\beta)2p^xJ_3\bigg\rbrack
\int dyy\chi_{_{P0}}^{(1)} \sqrt{y\over x}K_6\nonumber\\
& + & {2\over 3\pi}\bigg\lbrack
(x-\alpha\beta p^2+m_am_b)(J_1-4J_3) + (\alpha-\beta)
(J_2-4J_5)\bigg\rbrack p^2\nonumber\\
&\times&
\int dyy^2\chi^{(2)}_{_{P0}}\bigg\lbrack {4\over 3}K_7-K_1
\bigg\rbrack\ ,
\eeqn

where
\beq
\chi^{(0)}_{_{P1}} = -{(\alpha m_b + \beta m_a)J_1\over
J_1(x-\alpha\beta p^2+m_am_b)+J_2(\alpha-\beta)}
\ \chi^{(0)}_{_{P0}} \ ,
\eeq

\beq
\chi^{(0)}_{_{P2}} = {(m_a - m_b)J_1\over
J_1(x-\alpha\beta p^2+m_am_b)+J_2(\alpha-\beta)}
\ \chi^{(0)}_{_{P0}} \ ,
\eeq

\beq
\chi^{(0)}_{_{P3}} = {1\over 2}\ {J_1\over
J_1(x-\alpha\beta p^2+m_am_b)+J_2(\alpha-\beta)}
\ \chi^{(0)}_{_{P0}} \ ,
\eeq

\beq
\chi^{(1)}_{_{P0}} = {(x-\alpha\beta p^2+ m_am_b){2J_2\over xp^2}+
(\alpha-\beta)2J_3\over
J_1(x-\alpha\beta p^2+m_am_b)+J_2(\alpha-\beta)}
\ \chi^{(0)}_{_{P0}} \ ,
\eeq

\beq
\chi^{(2)}_{_{P0}} = {1\over xp^2}\ {(x-\alpha\beta p^2+ m_am_b)
{4J_3-J_1} + (\alpha-\beta)(4J_4-J_2)\over
J_1(x-\alpha\beta p^2+m_am_b)+J_2(\alpha-\beta)}
\ \chi^{(0)}_{_{P0}} \ .
\eeq

\bigskip

{\noindent {\it Scalar mesons}:}

\beqn
\Delta\chi_{_{S0}}^{(0)} &=&
\bigg\lbrack -{2\over 3\pi} (m_a+m_b)J_2  \int dy y
\chi_{_{S1}}^{(0)}(3K_1-2yK_3)\nonumber\\
&-&{2\over 3\pi}p^2(\alpha m_b-\beta m_a)J_1
\int dyy\chi_{_{S1}}^{(0)}(K_1+{2\over 3} yK_3)\nonumber\\
&-&{4\over 9\pi}p^2(\alpha m_b-\beta m_a)J_3
\int dyy \chi_{_{S1}}^{(0)}(3K_1-4yK_3)\bigg\rbrack\nonumber\\
&-& {2\over 3\pi}\bigg\lbrack (m_a+m_b)J_1+(\alpha m_b-\beta m_a)
{J_2\over x}\bigg\rbrack
\int dyy\chi_{_{S2}}^{(0)}\sqrt{yx}(3K_6-2\sqrt{xy}K_3)\nonumber\\
&-& {2\over 3\pi}p^2 (J_3-J_1)
 \int dyy \chi_{_{S3}}^{(0)}
\bigg\lbrack(2\sqrt{xy}K_6-{8\over 3}xyK_3)\bigg\rbrack\nonumber\\
&-& {2\over \pi}\bigg\lbrack (-x+\alpha\beta p^2+m_am_b)2J_2
-(\alpha-\beta)2p^2xJ_3\bigg\rbrack
\int dyy\chi_{_{S0}}^{(1)} \sqrt{y\over x}K_6\nonumber\\
& - & {2\over \pi}\bigg\lbrack
(-x+\alpha\beta p^2+m_am_b)J_1 - (\alpha-\beta)J_2\bigg\rbrack p^2\nonumber\\
&\times & \int dyy^2\chi^{(2)}_{_{S0}}\bigg\lbrack {4\over 3}K_7-K_1
\bigg\rbrack\ ,
\eeqn

where
\beq
\chi^{(0)}_{_{S1}} = -3(\alpha m_b - \beta m_a)J_1\ I_{S0}\ ,
\eeq

\beqn
\chi^{(0)}_{_{S2}} &=& - 3(m_a + m_b)J_1\ {2\over 3\pi}\ I_{S0}\nonumber\\
&+& (-x-\alpha\beta p^2 + m_am_b)\ {2\over 3\pi}\int dy y\chi^{(0)}_{_{S2}}
\lbrack 3\sqrt{y\over x}K_6 - 2yK_3\rbrack\ ,
\eeqn

\beq
\chi^{(0)}_{_{S3}} = {3\over 2}J_1\ I_{S0}\ ,
\eeq

\beq
\chi^{(1)}_{_{S0}} = -3\bigg[(-x+\alpha\beta p^2+ m_am_b){2J_2\over xp^2}-
 (\alpha-\beta)2J_3\bigg]\ I_{S0}\ ,
\eeq

\beq
\chi^{(2)}_{_{S0}} = -3\bigg[{1\over xp^2}\ (-x+\alpha\beta p^2+ m_am_b)
(4J_3-J_1) - (\alpha-\beta)(4J_4-J_2)\bigg] \ I_{S0}\ .
\eeq

\bigskip

\noindent $J_3$ and $J_4$ are given by,

\beqn
J_3 &=& {2\over \pi}\int\limits^\pi_0 d\theta{\sin^2\theta
\cos^2\theta\over  D(p^2,x,\cos\theta)}\nonumber\\
&=& {1\over A_aA_b} {1\over \beta c_1+\alpha c_3}
\bigg\lbrack {\alpha c_1\over D_1^2}+ {\beta c_3\over D_2^2}\bigg\rbrack
\eeqn

\beqn
J_4 &=& {2p\sqrt x\over \pi}\int\limits^\pi_0 d\theta{\sin^2\theta
\cos^3\theta\over  D(p^2,x,\cos\theta)}\nonumber\\
&=& {1\over 2A_aA_b} {1\over \beta c_1+\alpha c_3}
\bigg\lbrack {c_3^2\over D_2^2}- {c_1^2\over D_1^2}\bigg\rbrack
\eeqn

The kernels are defined by,
\beq
K_1(x,y) = {3\over 16\pi^2}\int d\theta \sin^2\theta\ G(x,y,\cos\theta)
\eeq

\beq
K_3(x,y) = {3\over 16\pi^2}\int d\theta {\sin^4\theta\over (x+y-2\sqrt{xy}
\cos\theta)}\ G(x,y,\cos\theta)
\eeq

\beq
K_4(x,y) = {3\over 16\pi^2}\int d\theta {\sin^4\theta\ \cos\theta
\over (x+y-2\sqrt{xy}
\cos\theta)}\ G(x,y,\cos\theta)
\eeq

\beq
K_6(x,y) = {3\over 16\pi^2}\int d\theta \sin^2\theta\ \cos\theta
\ G(x,y,\cos\theta)
\eeq

\beq
K_7(x,y) = {3\over 16\pi^2}\int d\theta \sin^4\theta\ G(x,y,\cos\theta)
\eeq

\beq
K_8(x,y) = {3\over 16\pi^2}\int d\theta {\sin^6\theta\over (x+y-2\sqrt{xy}
\cos\theta)}\ G(x,y,\cos\theta)
\eeq

\vfill
\eject


\begin{tabular}{|c|c|c|c|c|c|c|c|c|c|c|}
\hline
meson  & $\pi$ & $K^+$ & $s\bar s$ & $\eta_c$ &
$ D_s$ & $D_0$ & $\eta_B$ & $B_c$ &
 $B_s$  & $B_+$ \\
\hline
mass (MeV) & 135 & 494 &  & 2979 & 1969 & 1865 & & & & 5279\\
(exp.)& & & & & & & & & \\
\hline
mass (MeV) & 135 & 494 & 703 & 2821 & 1872 & 1756
& 9322 & 6126 & 5249 & 5149\\
\hline
$f_M$ (MeV) & 93 & 114 & & 213 & 148 & 125 & 287 & 207 & 119 & 119\\
\hline
$\alpha$ & 0.5 & 0.35 & 0.5 & 0.5 & 0.25 & 0.2 & 0.5 & 0.27 & 0.10 & .09\\
\hline
\end{tabular}
{\vglue 0.8cm\rightskip=3pc
\noindent
Table 1: Calculated pseudoscalar masses, corresponding center of mass
parameters $\alpha$ and decay constants $f_M$. Input parameters used for
all calculations were
the five quark bare masses and the two parameters $a$ and $\mu$
appearing in Eq. (6)}
\vglue 1.5cm

\begin{tabular}{|c|c|c|c|c|c|c|c|c|c|c|}
\hline
meson  & $\rho$ & $K^{*+}$ & $\phi$ & J/$\psi$ & $D_s^*$ & $D_0^*$ &
$\Upsilon $ & $B_c^*$ & $B_s^*$ & $B_+^*$ \\
\hline
mass (MeV) & 770 & 892 & 1019 & 3097 & 2110 & 2007 & 9460 &  & &5325\\
(exp.)& & & & & & & & & &\\
\hline
mass (MeV) & 760 & 976 & 1184 & 3100 & 2187 &  1997 & 9460 & 6277 &
5415 & 5290\\
\hline
$\alpha$ & 0.5 & 0.65 & 0.5 & 0.5 & 0.3 & 0.3 & 0.5 & 0.2 & .1 & 0.07\\
\hline
\end{tabular}
{\vglue 0.8cm\rightskip=3pc
\noindent
Table 2: Calculated vector meson masses and corresponding center of
mass parameters $\alpha$.}

\vglue 1.5cm

\begin{tabular}{|c|c|c|c|c|c|c|c|c|c|c|}
\hline
meson  & $u\bar u$ & $u\bar s$ & $s\bar s$ & $\chi_{c0}$ & $s\bar c$ &
$u\bar c$ &
$\chi_{b0} $ & $c\bar b$ & $s\bar b$ & $u\bar b$ \\
\hline
mass (MeV) &  &  &  & 3415 &  &  & 9860 &  & &\\
(exp.)& & & & & & & & & &\\
\hline
mass (MeV) & 593 & 903 & 1200 & 3605 & 2413 &  2127 & 10024 & 6733 &
5682 & 5462\\
\hline
$\alpha$ & 0.5 & 0.4 & 0.5 & 0.5 & 0.2 & 0.2 & 0.5 & 0.1 & 0.1 & 0.1\\
\hline
\end{tabular}
{\vglue 0.8cm\rightskip=3pc
\noindent
Table 3: Calculated scalar meson masses and corresponding center of
mass parameters $\alpha$.}
\vglue 0.8cm
\vfill
\eject

\begin{tabular}{|c|c|c|c|c|c|c|c|c|}
\hline
meson  & $\pi(2S)$ & $\rho(2S)$ & $K(2S)$ & $D_0(2S)$
 & $D_s(2S)$ & $D_s^*(2S)$ &  $\eta_c(2S)$ & $J/\psi(2S)$\\
\hline
mass (MeV) & 1300 & 1450 &  1460 & & & &  3415 & 3686\\
(exp.)& & & & & & & &  \\
\hline
mass (MeV) & 1368 & 1479 & 1580 & 2480  & 2600 &  2830 & 3470 & 3700
 \\
\hline
$\alpha$ & 0.5 & 0.5 & 0.5 & 0.25 & 0.25 & 0.3 & 0.5 & 0.5\\
\hline
\end{tabular}
\vglue 0.8cm

\begin{tabular}{|c|c|c|c|c|c|c|c|c|c|c|}
\hline
meson  & $B_0$ & $B_0^*$ & $B_s$ & $B_s^*$ & $B_c$ & $B_c^*$ &
 $\eta_B$ & $\eta_B$ & $\Upsilon$ & $\Upsilon$ \\
 & (2S) & (2S) &(2S) &(2S) &(2S) &(2S)  & (2S)&(3S) & (2S)& (3S) \\
\hline
mass (MeV)  & & & & & & & & & 10023 & 10355\\
(exp.)& & & & & &  & & & & \\
\hline
mass (MeV) & 5680 & 5740 & 5750 & 5770  & 6660 & 6800  & 9770 &
10150 & 9980 &  10300\\
\hline
$\alpha$ & 0.1 & 0.1 & 0.1 & 0.1 & 0.27 & 0.25 & 0.5 & 0.5 & 0.5 & 0.5\\
\hline
\end{tabular}
{\vglue 0.8cm\rightskip=3pc
\noindent
Table 4: Calculated excited states and corresponding center of
mass parameter $\alpha$.}

\vglue 1.5cm

\begin{tabular}{|c|c|c|c|c|c|c|c|}
\hline
meson  & $\pi^+-\pi^0$ & $K^0-K^+$ & $D_+-D_0$ & $B_0-B_+$
 & $K_0^*-K_+^*$ & $D_+^*-D_0^*$ & $B_0^*-B_+^*$ \\
\hline
mass (MeV) & 4.6  & 4.0 & 4.8 & 0.1 & 6.7 & 2.9 & \\
(exp.)& & & & & & & \\
\hline
mass (MeV) &3.2 & 4.0 & 3.3 & 1.7 & 5.8 &  4.6 & 3.4\\
\hline
\end{tabular}
{\vglue 0.8cm\rightskip=3pc
\noindent
Table 5: Calculated isospin and electromagnetic mass splittings.}

\vfill
\eject
\centerline {\bf Figure Captions}
\bigskip
\noindent Fig. 1.\ \ Configuration space potential obtained from equation (6).

\medskip

\noindent Fig. 2.\ \ Mass functions in units of $\Lambda_{QCD}$ as a function
of $Q^2/\Lambda^2_{QCD}$.

\medskip

\noindent Fig. 3.\ \ Normalized Bethe-Salpeter pseudoscalar wave functions
as functions of $Q^2/\Lambda^2_{QCD}$.

\medskip

\noindent Fig. 4.\ \ Bethe-Salpeter vector wave functions normalized to
one at $Q^2=0$.

\medskip

\noindent Fig. 5.\ \ Bethe-Salpeter scalar wave functions normalized to
one at $Q^2=0$.

\vfill
\eject

\end{document}